\documentclass[journal]{IEEEtran}
\ifCLASSINFOpdf
\else
\fi
\usepackage{graphicx}
\usepackage{dcolumn}
\usepackage{bm}

\newcommand{\ve}[1]{{\bf{#1}}}

\begin{document}
%
\title{On the use of the Pad\'{e}-Fourier approximation in fast evaluation of the Green's function of layered media}
%
%
%

\author{Yakir~Hadad,~\IEEEmembership{Senior Member,~IEEE,}
\thanks{Y. Hadad is with the School
of Electrical and Computer Engineering, Tel Aviv University, Tel Aviv, 69978 Israel, e-mail: hadady@eng.tau.ac.il}
\thanks{Manuscript received Nov 19, 2021.}}

\maketitle

\begin{abstract}
Efficient Green's function evaluation in layered media is a holy-grail of wave theory in general and for electromagnetics in particular. While there is a very large amount of knowledge in this context with vast literature, there are yet challenging cases such as the Green's function in thick lossy media and the Green's function at thick media with negative parameters. Here we propose a technique that can nicely tackle these issues. Our approach is based on a rational function approximation of the spectra using the Fourier-Pad\'{e} approximation that is carried out in a conformal mapped spectral plane. We show that this approach can be used in challenging scenarios such as very thick and lossy layers, materials with negative parameters such as in plasmonics, and even to approximate a dominant branch-cut contribution far from the source.
\end{abstract}


%
\IEEEpeerreviewmaketitle

\section{Introduction}
The Green's function (GF) encapsulates  a complete description of the wave dynamics in a given medium. It is also inherent to any numerical integral equation solver, and used in this context to express the fields at any point in the domain of solution in terms of the unknown source distribution. As such, it is impossible to stress enough the importance of the ability to fast evaluate the GF. A particular case of significance is the GF in layered media which is critical for many applications, including integrated electronic circuit solvers, modelling propagation in various electromagnetic and acoustic guiding and radiation systems, for computer aided design tools for RF and optical flatland devices, as well as for modelling of electromagnetic and elastic propagation in the earth for geosensing purposes, just to name a few.
%
Due to their clear importance, naturally, techniques for fast evaluation of the Green's function in layered media have become a holy-grail in wave theory with vast existing literature.

\begin{figure}[h]
\begin{center}
\noindent
  \includegraphics[width=0.5\textwidth]{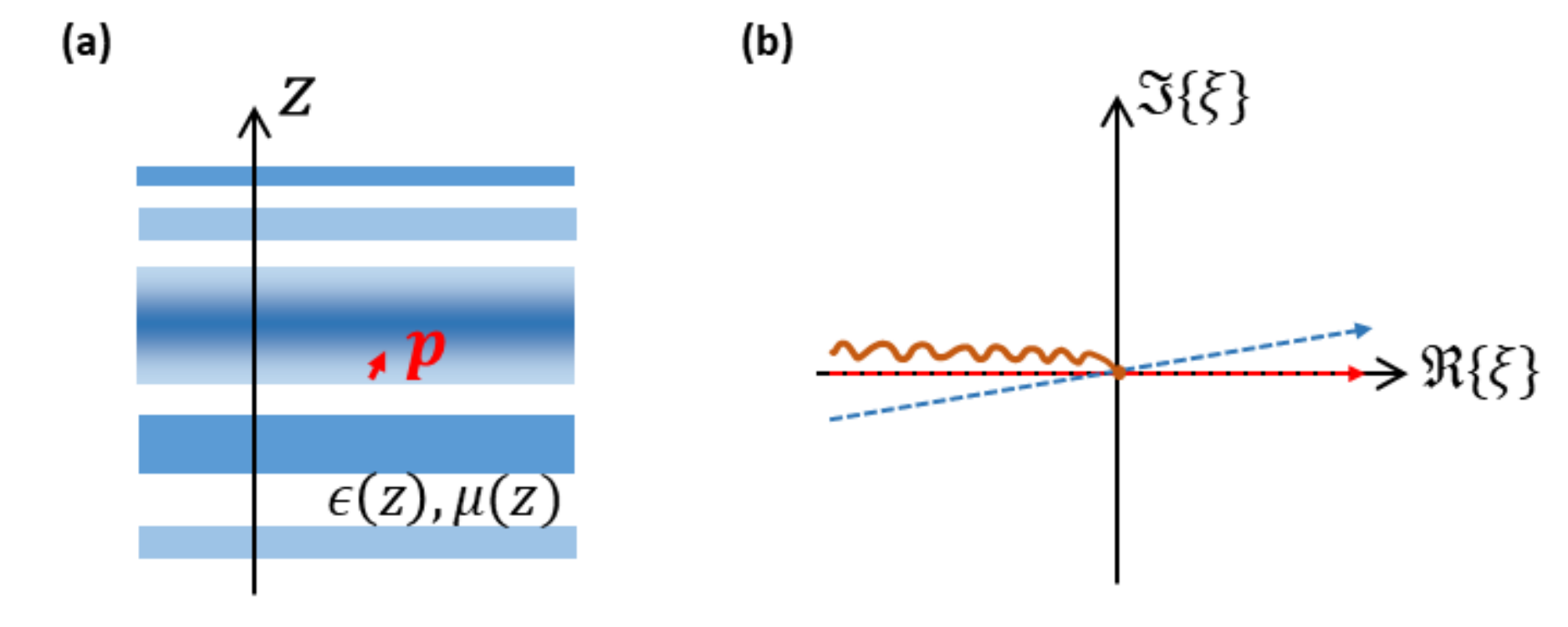}
  \caption{(a) Illustration of a plane stratified medium with electric dipolar source excitation. (b) The complex $\xi$ plane, and the Sommerfeld integration path on the real $\xi$ axis (brown). The wiggly line stands for the branch-cut of the Hankel function in Eq.~(\ref{Eq1}). The tilted dashed line representa the modified integration path to avoid pole singularities on the real $\xi$ plane.}\label{Fig1}
\end{center}
\end{figure}

The excitation response due to a dipolar source with moment $\ve{p}$, namely, the Green's function, in plane stratified media with layers normal to the $z$-axis  typically takes the following spectral  integral  form \cite{Felsen,Chew}
\begin{equation}\label{Eq1}
G(\ve{r})=\int_{-\infty}^\infty g(\xi;z,z')H_n^{(2)}(k\xi\rho)d\xi
\end{equation}
where the permittivity and permeability, $\epsilon(z)$ and $\mu(z)$, are piecewise continuous functions of $z$. The dipole source is located at $\ve{r}'=(0,0,z')$ and the observer at $\ve{r}=(x,y,z)$.   See Fig.~\ref{Fig1}(a) for illustration. In Eq.~(\ref{Eq1}), $G(\ve{r})$ stands for any of the field components ($x,y,$ or $z$, electric or magnetic field, as well as scalar or vector potentials), $\rho=\sqrt{x^2+y^2}$,  $\xi=k_t/k$ is the normalized transverse wavenumber, $k=\omega/c$,  where $c$ is the speed of light in vacuum,   and $g$ is the one-dimensional (1D) spectral Green's function that is a solution to a Sturm-Liouville (SL) problem which encapsulates completely the wave physics of the problem \cite{Felsen}. $H_n^{(2)}$ is the Hankel function of the second kind and order $n=0,1,2$, the order depends on the specific field component. Here and henceforth, and time dependence $e^{j\omega t}$ is assumed and suppressed. In Eq.~(\ref{Eq1}) the integration is taken along the so-called Sommerfeld integration path \cite{Felsen, Chew} on the complex $\xi$ plane, denoted by the red dashed line that is shown in Fig.~\ref{Fig1}(b). In the figure, the wiggly brown line denotes the branch cut of the Hankel function.

During the years many methods have been developed to evaluate the integral in Eq.~(\ref{Eq1}). Generally speaking, most of those can be associated with one of the following four categories.  ($i$) The semi-analytical approach in which one starts by finding the steepest descent path (SDP),  finding the singular points, i.e., poles and branch points, and finally deforming the integration path and integrating numerically along the SDP \cite{Felsen,Chew}. However, finding the SDP may be a tricky task, particularly in the case of multilayered media. Therefore, in the second approach ($ii$) the integration is carried out over a simpler path but using some smart integration quadrature role \cite{Chew, Simsek2006, Michalski1998, Michalski2016}. Nonetheless, in this case again, one has to have at least some prior knowledge regarding the singular points. ($iii$) The third approach departs from direct numerical integration.  Instead, in this approach it is suggested to replace the spectral GF kernel, namely the 1D Green’s function by an approximation using a set of complex exponents. Then, one can integrate analytically term by term over the expansion functions and get a quasi-closed form expression for the GF \cite{Fang1988, Aksun1992, Aksun1995,Aksun1996, Tokgoz2000, Ge2002, Shuley2002, Yuan2006}. Also in this approach, to make it accurate and efficient one has to extract first the poles and low frequency singularities \cite{Chow1991, Dural2005, Song2013}. Extracting the pole singularities can impose challenges  when considering thick and multilayered media that supports many modes, and especially when considering lossy materials as well as materials with negative permittivity and permeability, in which case the poles are no longer located on the real spectral axis but instead spread onto the complex plane. This brings us to the fourth category ($iv$) that avoids the need to have prior knowledge regarding the singular points. In this approach it is proposed to replace the 1D Green’s function by a rational function approximation \cite{Okhmatovski2002, Okhmatovski2004, Kourkoulos2006, Medina2007}. In this approach, instead, there are two main  issues. First, and more practical, is how to find the rational function approximation in an optimal way that balances between accuracy and numerical efficiency. Second,  and more fundamental, is related to the continuous spectrum contributions due to the branch cut singularities in the original 1D GF function. Can we trustfully replace them by an artificial discrete spectra in the form of additional finite set of poles? This  is a nontrivial point that has to be addressed carefully \cite{Mesa2008}.

The main contribution of this paper is in the proposal of the Pad\'{e}-Fourier approximation as a way to derive a rational function approximation that  nicely tackle the issue of proper description of the continuous  spectrum (lateral wave) using finite number of poles, and enables a rational function  approximation that is globally optimal in the sense that at infinity it convergence to the value of the original spectral function that is approximated. This cannot be achieve in the original spectral plane but can be done only after a conformal transformation of the original spectral plane to a Cayley transform plane.  The suggested approach is robust, and works with  uniform computational complexity, in various challenging cases such as of  thick lossy and lossless multilayered media, as well as media with negative parameters, i.e., primitivity and permeability, and in particular in open layered media in situations where the lateral wave due to a branch cut singularity, is dominant.

\section{Formulation of the fast method}

\subsection{The Pad\'{e} and the Pad\'{e}-Fourier approximations}\label{Pade}
The Pad\'{e} approximation is used similar to the Taylor series to approximate a given  function $f(x)$ about a certain point $x_0$. The derivation of the Pad\'{e} approximation starts by writing first the Taylor expansion up to certain order, $N$. Then, the Pad\'{e} approximation  which is a rational function with numerator and the denominator orders, $L$ and $M$, can be found. The coefficients of the Pad\'{e} polynomials are chosen  by writing down the Taylor expansion of the rational function and equating it to the Taylor expansion of the original function up to the same order $N$. Thus, locally, the Pad\'{e} and the Taylor approximations are identical. However, the Pad\'{e} approximation has an additional degree of freedom, that is, the difference between the orders of the numerator and denominator polynomials, $L-M$. This can be used to control the asymptotic behavior of the approximation, and thus to make it as similar as possible to that of the original function. This is why the Pad\'{e} approximation is typically considered as an efficient approximation with good \emph{global} accuracy also where the Taylor expansion on which it is based becomes highly inaccurate.
To demonstrate this behaviour we consider for example the approximation of the gaussian  function $f(x)=e^{-x^2}$ around the origin $x_0=0$.
Based on its Taylor expansion  $f_T(x)\approx 1-x^2+x^4/2-x^6/6$, we write two different Pad\'{e} approximations, $f_{P1}(x)\approx(1-x^2/3)/(1+2x^2/3+x^4/6)$ and $f_{P2}\approx1/(1+x^2+x^4/2)$.
In Fig.~\ref{Fig1} we compare the original function with the three approximations. We note that while the two Pad\'{e} approximations are based on the same Taylor expansion, in light of the extremely fast asymptotic decay of the gaussian function,  the second Pad\'{e} approximation $f_{P2}$ that decays the fastest  is the globally best.

\begin{figure}[h]
\begin{center}
\noindent
  \includegraphics[width=0.5\textwidth]{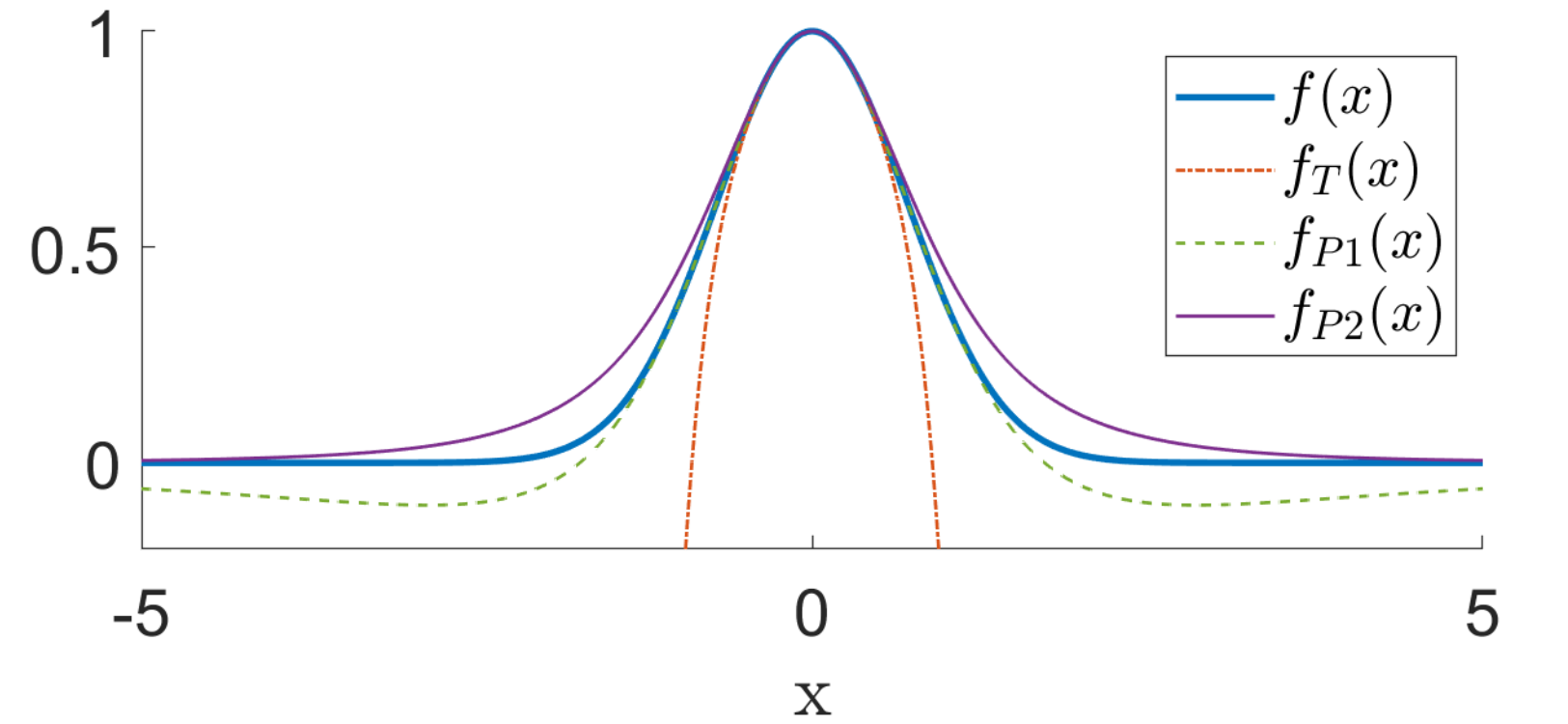}
  \caption{The Pad\'{e} approximation is globally better than the Taylor approximation provided that its asymptotic behaviour is properly chosen.}\label{Fig2}
\end{center}
\end{figure}

The Pad\'{e}-Fourier approximation is one of the generalizations of the Pad\'{e} approximation. The main difference between these two is the starting point. Namely, instead of using the Taylor expansion as the initial power series, in the Pad\'{e}-Fourier approximation, the Fourier expansion is used. The idea originates from approximation theory and harmonic analysis \cite{Pade1}, and demonstrated as an efficient tool for instance in the context of nonlinear dynamics where approximation for the solution of certain differential equations that give rise to shock waves is sought.  In this case, expressing the solution  using a mere Fourier expansion is not enough due to the Gibbs phenomenon that makes it impossible to adequately capture the discontinuity in the wave form. However it has been shown \cite{Pade2} that by having a ratio of two Fourier expansions it is possible to nicely describe a discontinuity of a function or its derivatives.

In our context, this hints that it may be possible to take advantage of this property of the Pad\'{e}-Fourier approximation also to capture the branch point and branch cut discontinuity in the complex plane of a singular complex function. This is the key idea of the proposed approach described below. We would like to approximate the 1D Green's function $g$ in Eq.~(\ref{Eq1}). The latter may or may not have a branch point singularity, depending whether the layered problem is ``closed'' or ``open''. In the former case, namely, the ``closed problem'', the SL operator is compact and therefore its spectrum is only discrete. In this case it is natural to seek for approximation of the 1D Green’s function by a rational function that exhibits only pole singularities. As opposed to that, we consider also a second type of problems, the ``open problem'', in which case the SL operator is non-compact and consists of discrete as well as continuous spectra in the form of branch point that introduces a finite `jump' in the complex function $g$.
Thus, the established ability of the Fourier-Pad\'{e} approach to approximate a `jump' discontinuity will become useful in our proposed method.


\subsection{Lossless versus lossy media}
The Sommerfeld path, shown in red-dashed line in Fig.~\ref{Fig1}(b) is the original  integration path  in Eq.~(\ref{Eq1}). It goes along the real-$\xi$ axis, slightly below the branch cut of the Hankel function when $\mbox{Re}\{\xi\}<0$. If the medium is lossy, all the mode poles drift out off the real-$\xi$ axis and therefore no pole singularities are present along the integration path. However, if the medium is lossless, all the pole singularities are located on the real-$\xi$ axis. By gently deforming the integration path this issue is resolved. A simple way of deformation is to define a new integration path using the variable $\mu$ that is related to $\xi$ via the mapping $\mu\mapsto\xi$
\begin{equation}\label{Eq2}
\xi=(1+j\alpha)\mu,\quad d\xi=(1+j\alpha)d\mu \quad \alpha\ll 1.
\end{equation}
Thus, an integration that is carried along the real-$\mu$ axis, between $-\infty$ and $\infty$, is mapped, in the $\xi$-plane, to a path that goes  along a tilted straight line with slope $\sim\alpha$ with respect to the $\mbox{Re}\{\xi\}$ axis as shown in the blue-dashed arrow in Fig.~\ref{Fig1}(b). In general, $\alpha>0$, typically taken in the range of $0-0.1$, where for lossy media we use lower values closed to zero. In the lossless medium examples that we study below we  use $\alpha\approx0.07$, while for the lossy medium examples we take $\alpha\approx0.02$. Note that in any case the solution should not depend on the particular choice of $\alpha$.

\subsection{Introducing an alternative spectral plane by the conformal Cayley transformation}
As a first step towards application of the Pad\'{e}-Fourier approximation we use the conformal Cayley transform to map the real-$\mu$ axis (i.e., the slightly tilted integration path in the $\xi$ plane) onto the unit circle. To that end we define a new spectral variable, $\eta$, by the mapping $\eta\mapsto\mu$
\begin{equation}\label{Eq3}
\mu=-j\frac{1+\eta}{1-\eta}\quad d\mu=-j\frac{d\eta}{(1-\eta)^2}.
\end{equation}
This is the place to emphasize that the use of the conformal mapping in Eq.~(\ref{Eq3}) introduces two major benefits. First, it provides a natural recipe for a nonuniform sampling of the entire real $\mu$ axis, in a manner that prefers the region near the origin that is more interesting spectrally, and leads to  a dilute sampling as $\mu\rightarrow\pm\infty$. Second, but most importantly, in the Cayley variable plane, $\eta$, it can be easily shown that the spectral green's function approaches a constant as $\eta\rightarrow\infty$ in any direction. As discussed in Sec.~\ref{Pade}, this property is of a very high significance in the evaluation of a Pad\'{e} approximation since it enables to control not only the local behaviour of the approximation, but also its asymptotic features, and turning it to be globally optimal. In contrast, the spectral green's function in the $\xi$ (or $\mu$) plane tends to zero, infinity, or it is highly oscillatory, depending on the direction in which $\xi$ (or $\mu$) approaches infinity. This point and its analytic consequences are further discussed below.

\begin{figure}[h]
\begin{center}
\noindent
  \includegraphics[width=0.5\textwidth]{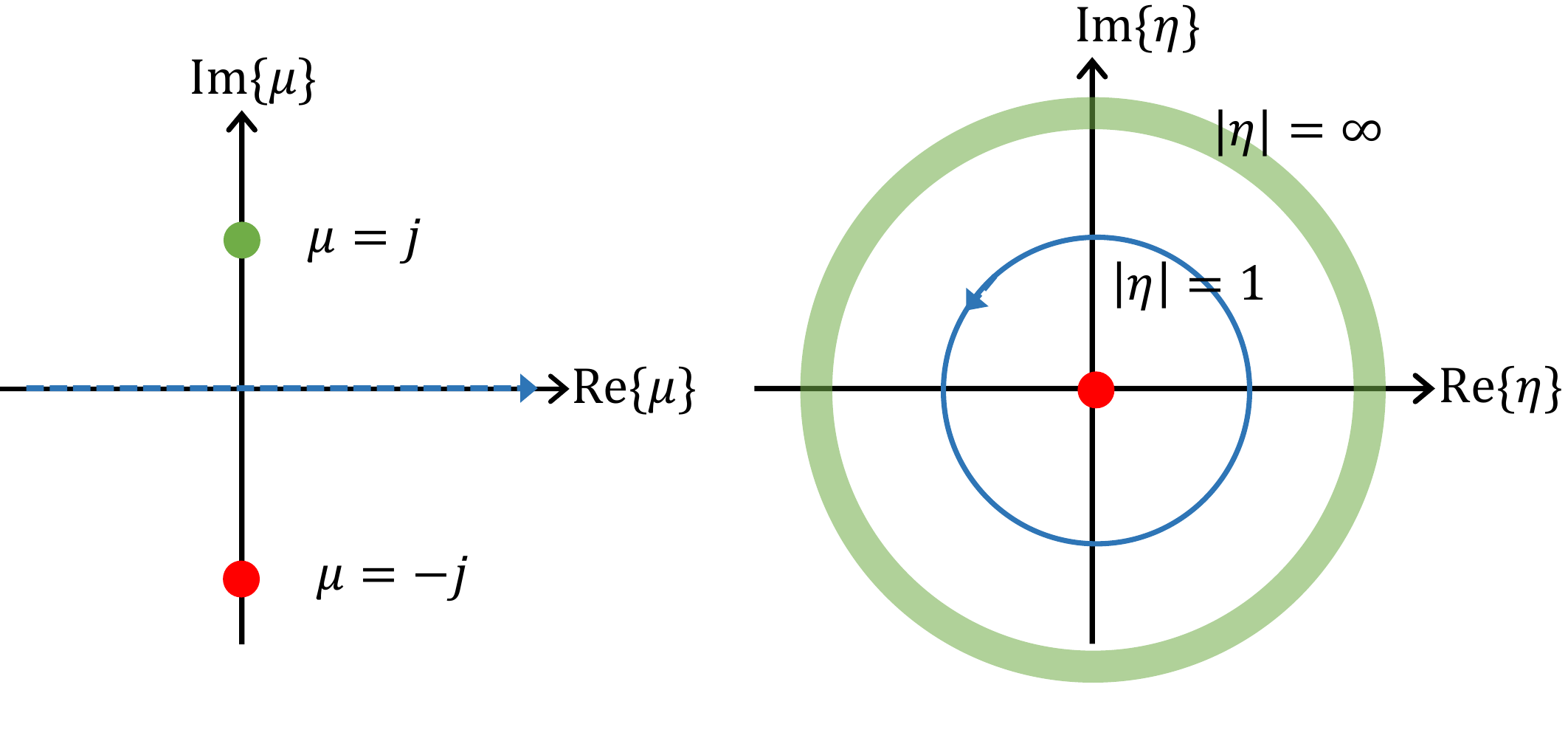}
  \caption{The complex $\mu$ plane is mapped onto the complex $\eta$ plane. The real $\mu$ axis is mapped to the unit circle, and $\mu=\pm j$ are mapped to $\eta\rightarrow0,\infty$. Hence, the asymptotes of $g(\eta;z,z')$ are constants implying that for the Pad\'{e}-Fourier approximation we should take $L=M$. }\label{Fig3}
\end{center}
\end{figure}

By using the Cayley transform, the spectral integral in Eq.~(\ref{Eq1}) is replaced by
\begin{eqnarray}\label{Eq4}
  G(\ve{r}) &=& \oint_{C1} f(\eta; z,z')H_n^{(2)}(k\xi\rho)d\eta\nonumber \\
  f(\eta;z,z') &=& \left(1+j\alpha\right)\frac{-j}{(1-\eta)^2}g(\eta;z,z')
\end{eqnarray}
In Eq.~(\ref{Eq4}), for brevity $f(\eta;z,z')$ stands for the kernel of the new spectral representation, $g(\eta;z,z')=g\left(\xi(\mu(\eta));z,z'\right)$,  and $\xi=\xi(\mu(\eta))$. We note that under this mapping $\mu = j\mapsto\eta=\infty$ and $\mu=-j\mapsto\eta=0$. See Fig.~\ref{Fig3} for illustration. Thus, the new kernel $g(\eta;z,z')$ tends asymptotically to \emph{constant values} at the two limits $\eta\rightarrow0$,  $|\eta|\rightarrow\infty$. We will recall this important observation in the following.

In light of Eq.~(\ref{Eq3}), in the complex $\eta$-plane the integration path  is mapped to the unit circle $C_1$ ($|\eta|=1$). Moreover, by the transformation in Eq.~(\ref{Eq2}) we enforce that along the integration path $g(\eta;z,z')$ has  no pole singularities.  Thus, we may define an annulus around the unit circle, $1-\delta<|\eta|<1+\delta$, on which the function $g(\eta;z,z')$ is analytic and therefore it is expressible in terms of a Laurent series with expansion coefficient $c_n$,
\begin{equation}\label{Eq5}
g(\eta;z,z') = \sum_{n=-\infty}^{\infty} c_n\eta^n \approx \sum_{n=-N}^{N} c_n\eta^n.
\end{equation}
Specifically on the unit disk, $\eta=e^{j\phi}$, $\phi\in[0,2\pi)$ and the Laurent series turns to be  nothing but a Fourier expansion that its coefficients can be found using a standard and efficient Fast Fourier Transform (FFT). Once the coefficients $c_n$ are known, the Pad\'{e}-Fourier approximation is found by equating the truncated Laurent series up to order $N$ to a rational function with polynomials of order $2L$ and $2M$, in the numerator and the denominator, respectively. Thus,
\begin{equation}\label{Eq6}
\sum_{n=-N}^{N} c_n\eta^n = \frac{P_L(\eta)}{Q_M(\eta)}=\frac{\sum_{l=-L}^{L}a_l\eta^l}{\sum_{m=-M}^{M}b_m\eta^m}
\end{equation}
Now, recall  that  at $|\eta|\rightarrow\infty$ and $\eta\rightarrow0$ the function $g(\eta;z,z')$ tends to  constant values.  This implies   that necessarily $L=M$, and specifically that $a_{-N}/b_{-N} = g(\xi(\mu=j),z,z')$, and $a_N/b_N = g(\xi(\mu=-j),z,z')$. Using Eq.~(\ref{Eq5}) a linear system for the unknown coefficients can be obtained \cite{Pade2} (see Eq.~(2.10) there).
Once solved, we can approximate the Green's function by
\begin{equation}\label{Eq7}
G(\ve{r})\approx\oint_{C1} \frac{-j(1+j\alpha)}{(1-\eta)^2}\frac{P_M(\eta)}{Q_M(\eta)} H_n^{(2)}(k\xi\rho)d\eta.
\end{equation}
In the integration over $C_1$ the singular point at $\eta=1$ that is an artifact of the conformal transformation in Eq.~(\ref{Eq3}) should be excluded. Then, the integration over the unit circle can  be replaced with a summation over the residues of the poles due to the rational function approximation. Thus, the rather involved task of searching complex poles of a general function $g(\xi;z,z')$ has been significantly simplified to the task of searching the zeros of the polynomial function $Q_M(\eta)$. Let us denote the set of these zeros by $\{\eta_z\}$, then Eq.~(\ref{Eq6}) will be replaced by
\begin{equation}
G(\ve{r})\approx 2\pi j \! \!\!\sum_{|\eta_z| < 1}\!\frac{-j(1+j\alpha)}{(1-\eta_z)^2} \frac{P_L(\eta_z)}{\partial_\eta\left[Q_M(\eta_z)\right]} H_n^{(2)}(k\xi_z\rho)
\end{equation}
where the summation is taken only over the poles that are located inside the unit circle, and $\xi_z=\xi(\mu(\eta_z))$.

Up to here we discussed the analytical details of the proposed fast approach. The argument is that the Pad\'{e}-Fourier expansion is capable of describing discontinuities and hence it has a potential to nicely capture the branch-point contribution to the Green's function. Furthermore, the Pad\'{e}-Fourier approximation takes into account not only the dynamics of the complex function near the region of singular points but also, in light of the conformal mapping that we carried out, we showed that the choice $L=M$ also captures the dynamics at infinity and the origin. In that sense, the approximation we propose is \emph{globally optimal}.
In the following section we demonstrate the strength of the proposed algorithm through various  examples.

\section{Examples}
In the following we consider a few examples that demonstrate various aspects of the proposed technique. We discuss the concept of true and artificial poles, and show different rules of the latter as a means to describe the near fields to the source, as well as to describe  the lateral wave that is excited due to the continuous spectrum contribution in open structures.
\subsection{True and artificial poles}
\subsubsection{Lossy parallel plate waveguide}
\begin{figure}[h]
\begin{center}
\noindent
  \includegraphics[width=0.45\textwidth]{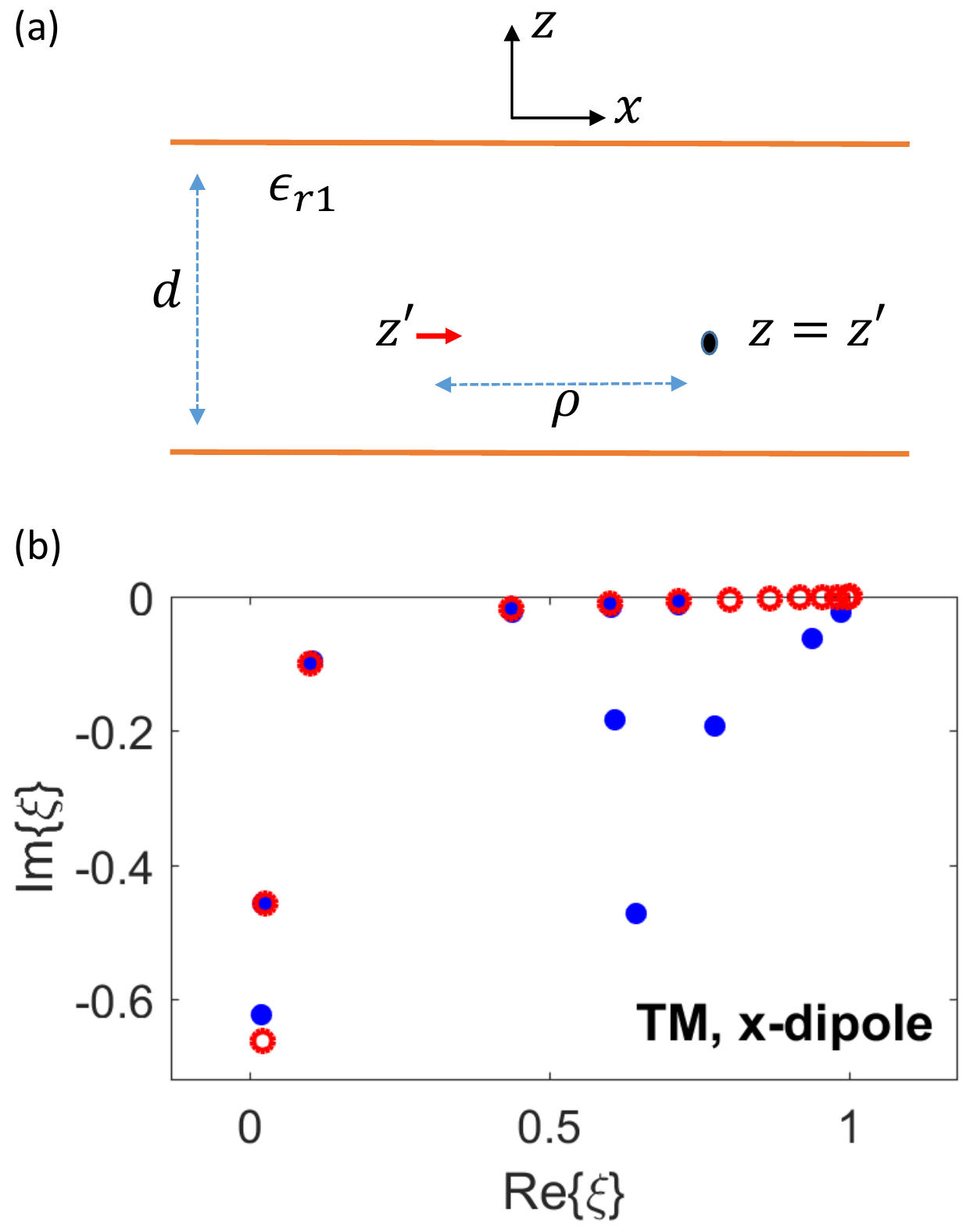}
  \caption{(a) A thick parallel plate waveguide filled with lossy dielectric. The $x$ polarized dipole source, and the observer are located on the same plane $z=z'$. (b) Red circles: actual modes of the structure calculated analytically. Blue dots: modes obtained through our formalism in this paper. Some of them are 'true' while other are 'artificial'.}\label{Fig4}
\end{center}
\end{figure}

Consider a thick parallel plate waveguide that consists of two parallel PEC plates located at distance  $d=5\lambda_0$ of each other. The waveguide is filled with dielectric material with relative permittivity $\epsilon_{r1}=1-j0.05$, as shown in Fig.~\ref{Fig4}(a). In this example the 1D green’s function for the field's $x$-component of the TM fields due to a $x$-polarized dipole reads
\begin{eqnarray}
g(\xi;z,z') &=& -\frac{Z_{TM}/2}{1-\exp(-2jk_z d)}\times\left(e^{-jk_z z_1}\right. \nonumber \\
&-& \left. e^{-jk_z z_2} - e^{-jk_z z_3} + e^{-jk_z z_4} \right)
\end{eqnarray}
where $Z_{TM}=Z_0\sqrt{\epsilon_{r1}-\xi^2}/\epsilon$, $Z_0=120\pi\Omega$, and $z_1=|z-z'|$, $z_2=z+z'$, $z_3=2d-(z+z')$, $z_4=2d-|z-z'|$. This 1D green's function contains only pole singularities located at
\begin{equation}
\xi_{p,n}=\sqrt{1-\frac{1}{\epsilon_{r1}}\left(\frac{n\pi}{k_0d} \right)^2}
\end{equation}
where $k_0=2\pi/\lambda_0$ is the wavenumber of vacuum, and $z'$ [$z$] denotes the location of the source [observer]. Specifically, for the numerical example that is discussed next we use $z=z'=0.1\lambda_0$.

These poles are shown in Fig.~\ref{Fig4}(b) by the red circles. When we use our method with $L=M=20$, we search for the zeros of $Q_M(\eta)=0$ and transform them using Eq.~(\ref{Eq3}) to the $\xi$ plane we get another set of poles that are denoted by blue dots. Some of these poles correspond to true modes of the structure, that are excited in this particular source setup. These are the blue dots that coincide with the red circles in the Fig.~\ref{Fig4}(b). In addition, the red circles that do not coincide with blue dots correspond to modes the structure, that are not excited at this particular excitation setup, i.e., source type, location, and orientation.  However, we also see isolated blue dots, these correspond to artificial poles that can be associated with  \emph{artificial} modes of the system, that their role is merely to take care of the \emph{near fields} in the vicinity of the source and therefore are \emph{necessarily evanescent}.

\subsubsection{Plasmonic half-space}
Let us stress  the last point regarding the role of the artificial poles further, and consider this time an open plasmonic structure. The upper space is vacuum ($\epsilon_{r1}=1$) and the lower space consists of a plasmonic material with $\epsilon_{r2} = -1.1 -0.1j$. Here, only one mode, of a TM type is supported, that is the surface plasmon polariton wave that propagates at the interface between the two half spaces. Its wavenumber is given analytically by
\begin{equation}\label{Eq_SPP}
\xi_{\mbox{\scriptsize SPP}}=\sqrt{\frac{\epsilon_{r1}\epsilon_{r2}}{\epsilon_{r1}+\epsilon_{r2}}}.
\end{equation}
With the parameters assumed above, this yields $\xi_{\mbox{\scriptsize SPP}}=2.63-j0.65$.
On the other hand, following our formalism in this paper, with a source located inside the plasmonic material at $z'=-0.3\lambda_0$, and observer located in the vacuum at $z=-0.2$, and  with $L=M=10$, we find a set of poles. One of these is exactly placed at the location of the true pole of the system, but the other poles are merely used to describe the near evanescent fields at the vicinity of the source, as in the previous example, \emph{but also} to approximate  additional radiation waves that stem from the branch cut singularity that was absent in the electromagnetically closed parallel plate problem that was discussed before. The numerical distinction between the two types of artificial poles can be  done by considering their amount of  attenuation ($\mbox{Im}\{\xi\}$). The low loss artificial poles are intuitively associated with a manifestation of the radiation, while the high loss artificial poles are these that take care of the near fields.
\begin{figure}[h]
\begin{center}
\noindent
  \includegraphics[width=0.45\textwidth]{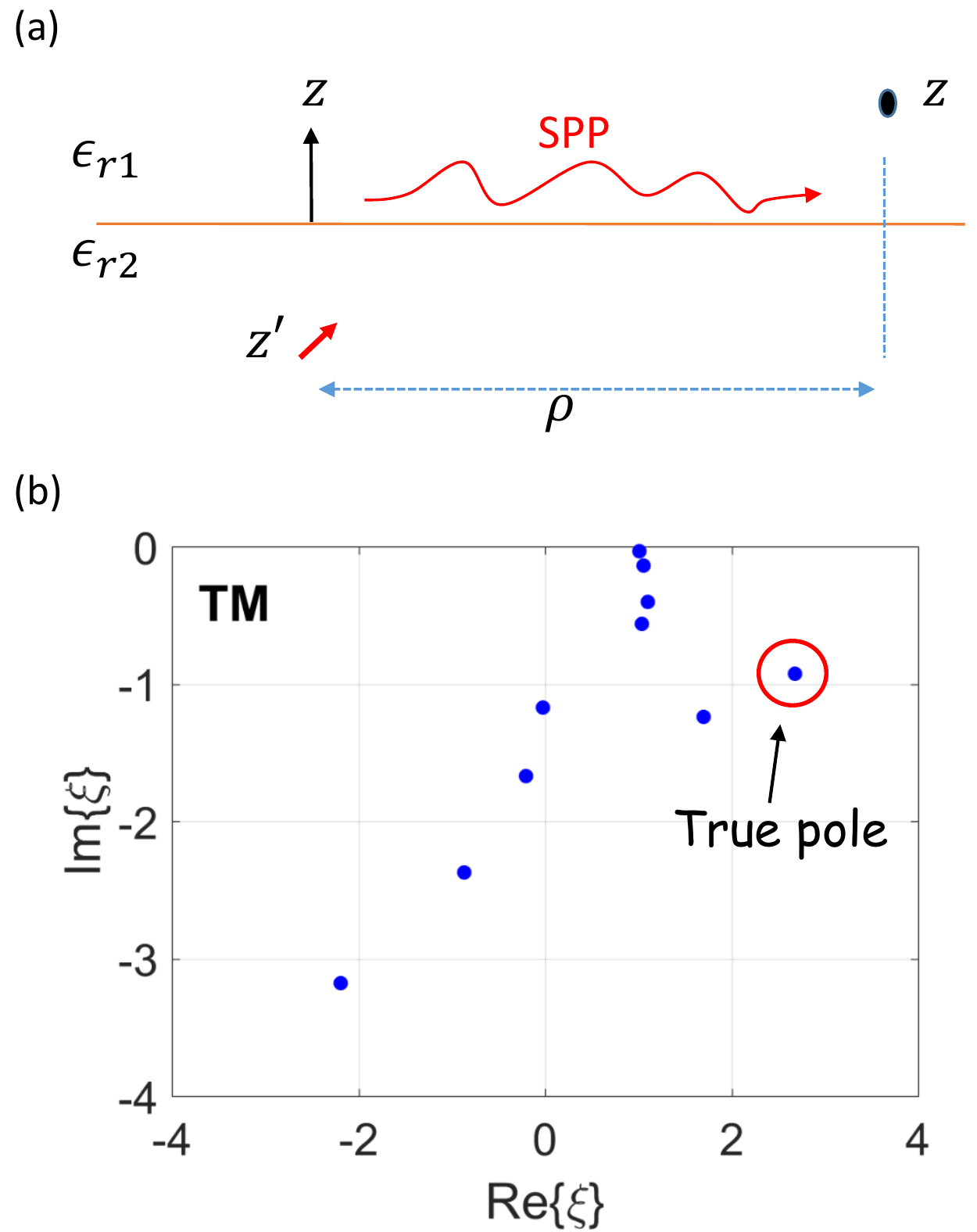}
  \caption{(a) An interface between plasmonic half-space and vacuum. (b)  Blue dots: modes obtained through our formalism in this paper. Only the one encircled is a true pole of the structure, the others are  'artificial' poles used to approximate the near field as well as the continuous spectrum that gives rise to radiation into the free space (this corresponds mathematically to a branch point singularity in the spectral green's function).}\label{Fig5}
\end{center}
\end{figure}

\subsection{Thick lossless and lossy dielectric waveguide}
\begin{figure}[h]
\begin{center}
\noindent
  \includegraphics[width=0.49\textwidth]{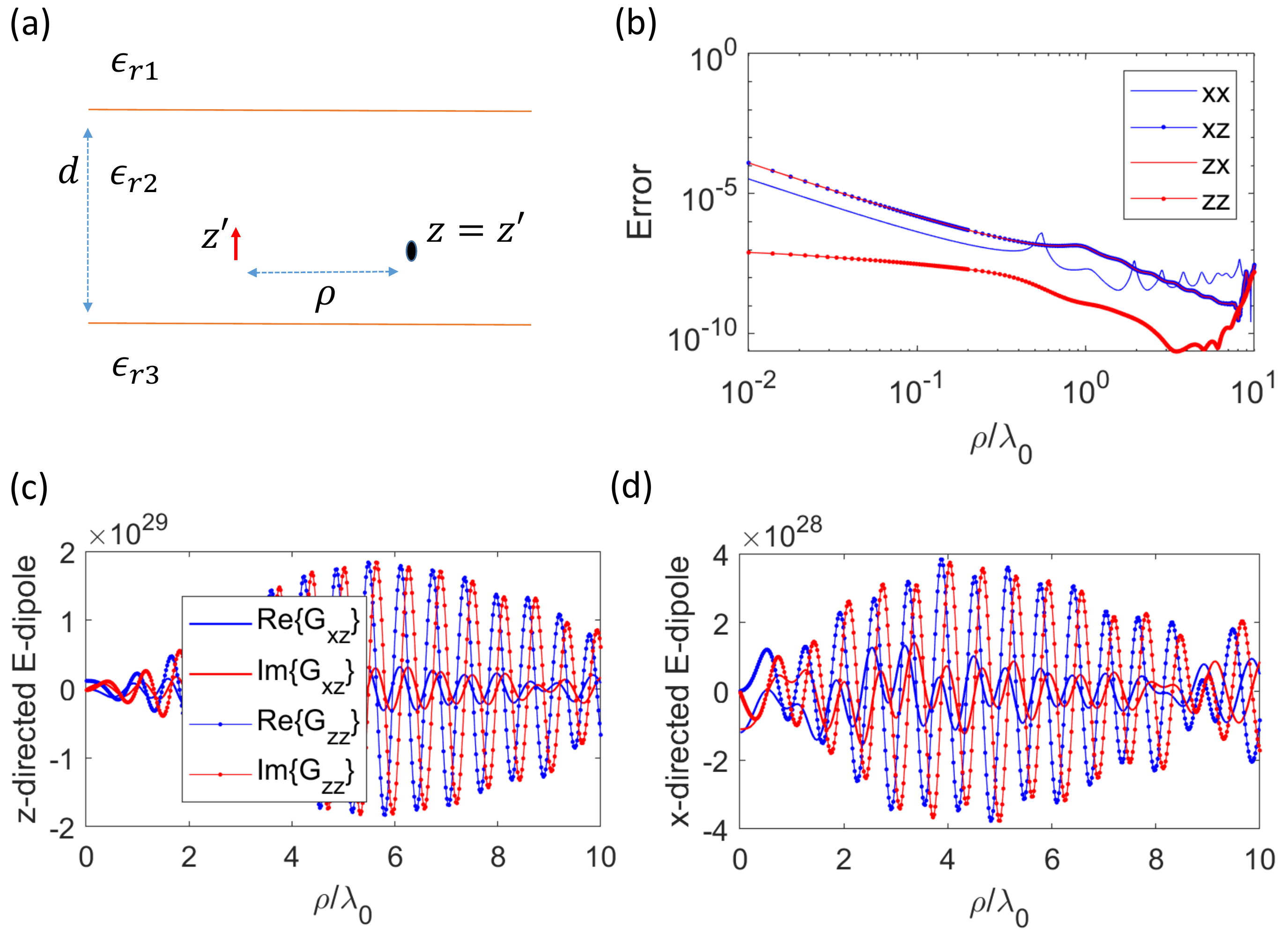}
  \caption{(a) A \emph{lossless} dielectric layer sandwiched between two half-spaces (b)  The relative error between our calculation and a brute-force numerically exact evaluation of the spectral integral. The typical errors is about $10^{-5}$ or less for long range of source-observer radial distance. (c) and (d) Various components of the electric dyadic Green's function, (c) for a $z$-polarized electric dipole inside the later. The observation point is located such that $z=z'=0.5\lambda_0$. (d) same as (c) for a $x$-polarized electric dipole.}\label{Fig6}
\end{center}
\end{figure}

In the following example we consider a thick dielectric waveguide with $\epsilon_{r1}=1, \epsilon_{r2}=2.79, \epsilon_{r3}=2.25$, the waveguide thickness is $d=6\lambda_0$, and it supports 16 TM modes and additional 16 TE modes. The source and the observer are located at $z=z'=0.5\lambda_0$. The structure is shown in Fig.~\ref{Fig6}(a).
The expression for the  spectral Green's function of the various electric Green's function dyadic is taken from a standard textbook \cite{Chew}, and are omitted here for brevity. Here, we focus on the  numerical evaluation  algorithm that is proposed in this paper.  Since the source and the observer are located at the same layer, and since the observer may be close to the source, we decompose the spectral Green's function into two components $g(\xi;z,z')=g_{{p}}(\xi;z,z')+g_{{s}}(\xi;z,z')$. Primary $g_{{p}}(\xi;z,z')$ and secondary $g_{{s}}(\xi;z,z')$, the former corresponds to the radiation in a uniform infinite medium with the parameter of the layer, and the second, is equal to the total spectral green's function after the substraction of the primary part. The primary part can then be evaluated analytically using the usual expressions for the space-domain green's function in homogenous medium. Here, we perform the numerical evaluation only for the secondary part of the spectral Green's function which is numerically challenging.   We compare our calculation to a brute-force ``numerically exact'' calculation and find that the relative error to be smaller then $10^{-5}$ for large range of distances between the source and the observer as shown in Fig.~\ref{Fig6}(b).  This accuracy is achieved using only $2L=100$ poles (true and artificial) in total, that are calculated using a standard zeros search algorithm for polynomial functions. In this simulation we used $\delta_1=0.07$.
After a distance of about $20\lambda_0-30\lambda_0$ the error substantially increases due to a too low accuracy in finding the true poles of the guided waves (recall that we do not perform the pole search on the original spectral green's function, but on its rational function approximation). This issue may be readily fixed by applying pole search refinement for the poles of the original spectral green's function. However, in this electrically long range it would typically be better to switch to and use the asymptotically evaluated solutions. Therefore, we do not care about this issue here anymore.  Fig.~\ref{Fig6}(c) and (d) show various components of the electrical field, for excitation by vertical and horizontal dipole sources, respectively.

\begin{figure}[h]
\begin{center}
\noindent
  \includegraphics[width=0.49\textwidth]{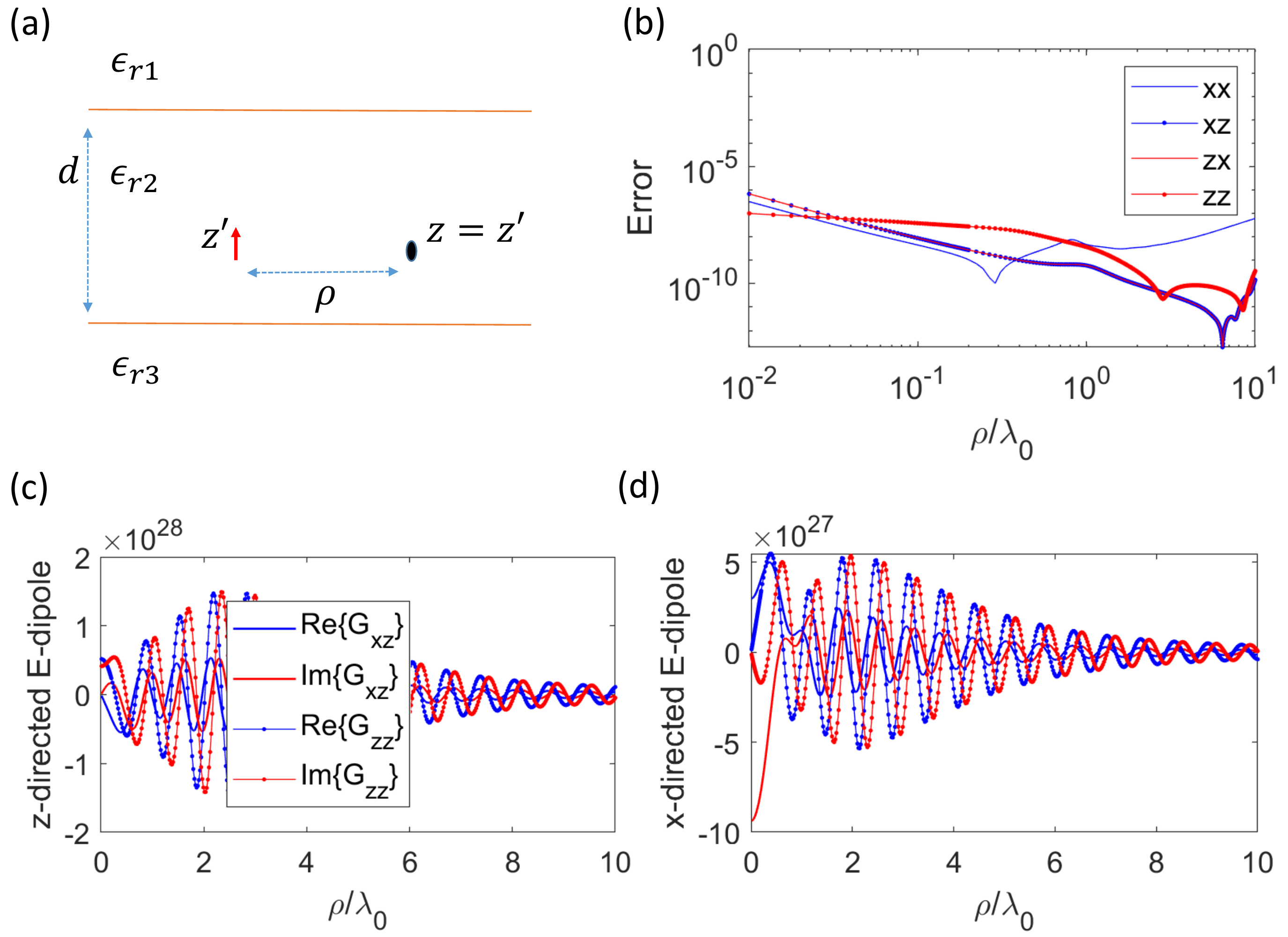}
  \caption{(a) A \emph{lossy} dielectric layer sandwiched between two half-spaces (b)  The relative error between our calculation and a brute-force numerically exact evaluation of the spectral integral. The typical error is about $10^{-5}$ or less for long range of source-observer radial distance. (c) and (d) Various components of the electric dyadic Green's function, (c) for a $z$-polarized electric dipole inside the later. The observation point is located such that $z=z'=0.5\lambda_0$. (d) same as (c) for a $x$-polarized electric dipole. }\label{Fig7}
\end{center}
\end{figure}

As soon as we introduce loss to the center layer, with $\epsilon_{r2}=2.79-j0.5$, the poles of the original spectral Green’s function spread into the complex plane, however \emph{the computational complexity of our calculation remains essentially unaffected} with $2L=100$ poles, and with $\delta_1=0.02$. The structure, is shown in Fig.~\ref{Fig7}(a), identical to Fig.~\ref{Fig6}(a) (except for the addition of loss). The relative error  compare with the numerically exact brute-force calculation are given  in Fig.~\ref{Fig7}(b), while Fig.~\ref{Fig7}(c) and (d) show the Green’s function components for the two excitation polarizations.  Remarkably,   also the accuracy remains as in the lossless case.

\subsection{Approximation of a lateral wave by a finite set of poles}
\begin{figure}[t]
\begin{center}
\noindent
  \includegraphics[width=0.49\textwidth]{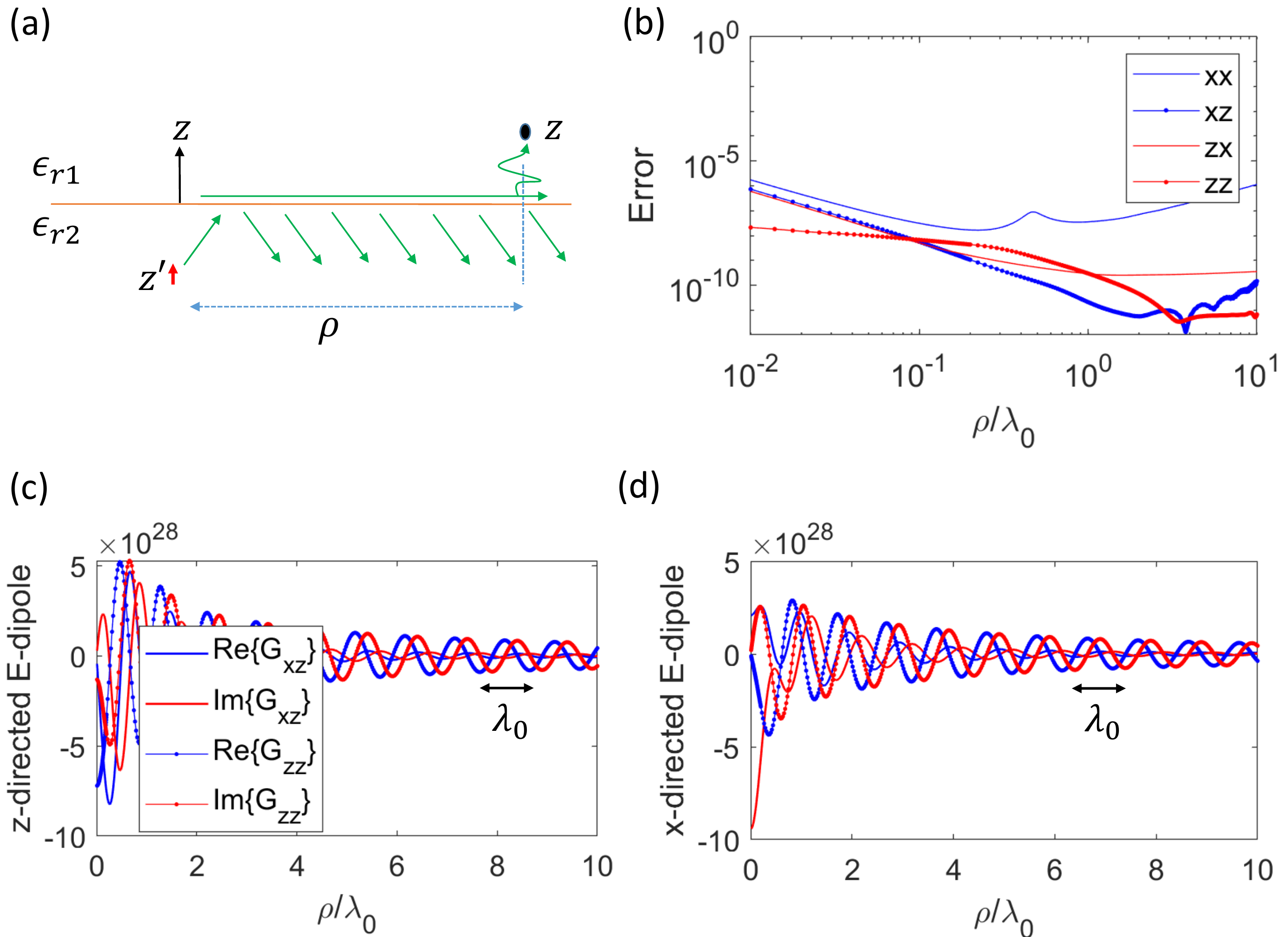}
  \caption{(a) A dielectric half space, and the propagation ray trajectory that corresponds to a lateral wave propagation. (b)  The relative error between our calculation and a brute-force numerically exact evaluation of the spectral integral. The typical errors is about $10^{-5}$ or less for long range of source-observer radial distance. (c) and (d) Various components of the electric dyadic Green's function, (c) for a $z$-polarized electric dipole inside the later. The propagation wavelength is shown to be equal to the free space wavelength, as should be in a lateral wave propagation. (d) same as (c) for a $x$-polarized electric dipole.}\label{Fig8}
\end{center}
\end{figure}

Lastly, we show that our approach is capable of nicely describing the branch cut singularity that corresponds to continuous wave radiation of spherical waves outside of the layered medium. To that end we design a numerical setup that consists of a dielectric half-space. The upper side is assumed to be vacuum while the lower side is considered denser, and heavily lossy, specifically we take $\epsilon_r=2.79-0.7j$. The source and the observer are located bellow and above the interface at $z'=-0.3\lambda_0$ and $z=0.2\lambda_0$, respectively, both are electrically close to the interface. In this configuration there are no guided or leaky modes since the spectral green's function involves no pole singularities. In addition, a saddle point contribution to the asymptotic evaluation of the spectral integral, that corresponds to a refracted ray from the source to the observer, will contribute weakly since  it will decay exponentially due to the loss in the lower half-space. Thus, \emph{from the asymptotic perspective the only contribution to the Green’s function will be a lateral wave} that corresponds to a ray that leaves the source, hits the interface at the critical angle, and then propagates parallel to the interface, \emph{just above the interface}, and thus experiencing no loss until it evanescently reaches the observer. This propagation process is illustrated in Fig.~\ref{Fig8}(a).
Hence, in order to evaluate the spectral integral correctly in this particular propagation setup we have to be able to properly estimate the contribution of the branch cut singularity.

We use our formalism to calculate the Green's function in this case, with $2L=100$ poles, and $\delta_1=0.02$, as used in the previous examples. First, as seen by Fig.~\ref{Fig8}(b), a relative error calculation in this case yields the same order of magnitude of errors, about $10^{-5}$ or less, as we had in the previous examples where the modal contribution was dominant. Moreover, as evident by Fig.~\ref{Fig8}(c) and (d) that show the excited field as a function of the radial distance,  the solution spatial periodicity becomes that of the free space when the radial distance between the source and the observer becomes large on the wavelength, so that the asymptotic picture of a propagating lateral wave becomes valid. Therefore, we can conclude that using our approach we are able to nicely capture the continuous spectrum waves that are associated with branch cut singularities, using a finite, moderate-size, set of poles. This behaviour is possible due to the ability of the Fourier-Pad\'{e} approximation to nicely approximate 'jump' discontinuities in the complex spectral plane, and in light of its ability to properly model the spectral green's function at infinity, as discussed in Sec.~\ref{Pade}.

\section{Computational complexity assessment}
As demonstrated in the examples above, the computational complexity of the suggested technique is nearly independent of the material parameters that may be lossless, lossy, and with negative or positive real parts, and number of layers. The computational complexity per frequency and per $z,z’$ pair is essentially that of 1 FFT calculation, 1 moderate-side  matrix (rank of 20-100) inversion, and polynomial zero search that is done efficiency using a standard algorithm such as MATLAB `\emph{roots}'.

\section{Conclusions}
To conclude, in this paper we proposed the Fourier-Pad\'{e} approximation as a tool to derive a rational function approximation that nicely captures the discontinuity in the complex plane due to the branch cut singularity of the 1D green’s function in open structures. To achieve this representation we applied a conformal mapping, specifically the Cayley transform,  from the original spectral plane $\xi$ to a `ring-like' plane $\eta$. By doing so, we showed that the spectral green's function at infinity $\eta\rightarrow\infty$ tends to a constant in any `direction of infinity'  which is a desired property of a function that is to be approximated by a Pad\'{e} type approximation. This way, the rational function approximation of the spectral green's function in the complex $\eta$ plane can be globally optimal. The Pad\'{e} approximation itself is carried out using a Lourent series that is found by analytic continuation of the spectral green's function on the $|\eta|=1$ ring. There, the Lourent series essentially coincides with a Fourier series on the function, on a $2\pi$ interval, that can be effectively evaluated using FFT.
We demonstrate the accuracy of the method in various challenging scenarios for Green's function evaluation, including thick layers, lossy layers, and negative index layers. And specifically we demonstrate its ability to approximate continuous spectrum waves (lateral waves) using a moderate number of poles. The computational complexity is finally discussed.


%

%
%

\section*{Acknowledgment}
This research was supported by the Israel Science Foundation (grant No. 1353/19), and by the Alon fellowship. The author would like to thank Dr. Yaniv Brick, Prof. Ben Z. Steinberg, and Prof. Ehud Heyman for useful discussions and critical reading of the manuscript.

\begin{IEEEbiography}{Yakir Hadad}
(Senior Member, IEEE) received
the B.Sc. and M.Sc. degrees (summa cum laude)
in electrical and computer engineering from Ben
Gurion University of the Negev, Be’er Sheva, Israel,
in 2006 and 2008, respectively, and the Ph.D. degree
in physical electronics from Tel Aviv University, Tel
Aviv, Israel, in 2014.
Between 2015 and 2017, he was a Post-Doctorate
Fellow with the Department of Electrical and Computer Engineering, The University of Texas at
Austin, Austin, TX, USA. During recent years, he
also spent several periods as a Visiting Scientist with the FOM Institute
Atomic and Molecular Physics (AMOLF), Amsterdam, The Netherlands,
in Fall 2015, and the University of Pennsylvania, Philadelphia, PA, USA,
in Spring 2013 and in Summer 2018. In 2017, he joined the Department of
Physical Electronics, Faculty of Engineering, Tel Aviv University, where he is
currently a Senior Lecturer (Assistant Professor). His research interest spreads
on a wide range of wave modeling problems as well as on analytical and
semi-analytical methods in electromagnetics and acoustics, with a particular
emphasize on wave phenomena in complex media with applications in
overcoming bounds of wave theory.
Dr. Hadad received the Felsen Award for Excellence in Electrodynamics
from the European Association for Antennas and Propagation in 2016, 
the 2017 recipient of the prestigious Alon Fellowship for Outstanding Young
Faculty from the Israeli Council of Higher Education, and the 2020 Krill Prize for excellence in research by the Wolf foundation.
\end{IEEEbiography}





\end{document}